\documentstyle[twoside,fleqn,espcrc2,epsfig]{article}
\title{
\vskip-3.cm
{\baselineskip14pt
\centerline{\normalsize\hfill DTP/97/76, September 1997}
\centerline{\normalsize\hfill hep-ph/9709377}}
\vskip22.cm
{\baselineskip10pt
\begin{flushleft}
{\small\em To appear in the Proceedings of the 
High Energy Physics International Euroconference on Quantum
Chromodynamics, 3--9 July 1997, Montpellier, France.}
\end{flushleft}}
\vskip-21.2cm
Improved QCD expansion and the problem of determination of
$\alpha_{s}$ and the condensates from the $\tau$ decays}

\author{
P. A. R\c{a}czka\address{
Centre for Particle Theory, University of Durham, South  Road, Durham
DH1~3LE, Great Britain 
}\thanks{On leave from Institute of Theoretical Physics, Warsaw
University, Warsaw, Poland}
}

\begin{document}

\begin{abstract}
The problem of determination of $\alpha_{s}$ and the condensates from
the moments of invariant mass distribution in the semileptonic $\tau$
decays is considered. It is investigated how the extracted values of
$\alpha_{s}(m^{2}_{\tau})$ and the condensates are affected by the the
renormalization scheme dependence of the next-to-next-to-leading order
perturbative corrections to the spectral moments.  A simplified
approach is used, in which the nonperturbative contributions are
approximated by the terms of dimension six in the SVZ expansion, which
arise from the four-quark condensates.
\end{abstract}

\maketitle

Recently there has been considerable interest in the determination of
the QCD parameters from the invariant mass distribution in the
semileptonic $\tau$ decays~\cite{first,ledi92b,exp,hoeck96}. It
appears that using specific 
moments of the invariant mass distribution one may obtain rather tight
experimental constraints for $\alpha_{s}$ and the QCD condensates. The
perturbative QCD corrections to the spectral moments --- which are
known to next-to-next-to-leading order (NNLO) --- play central role in
such a determination. These corrections are usually evaluated in the
$\overline{MS}$ renormalization scheme. However, in NNLO there is a
two-parameter freedom in the choice of the renormalization scheme
(RS). The numerical values of the perturbative corrections depend on
the choice of the scheme. It is therefore an interesting question, how
the values of the QCD parameters, extracted from the $\tau$ decay, may
be affected, if a different choice of the RS is made, motivated for
example by the principle of minimal sensitivity (PMS)
\cite{stev81}.

The theoretical framework adopted in the analysis of the $\tau$ decay
data involves the $R^{kl}_{\tau,V/A}$ moments, defined by the
relation~\cite{ledi92b}: 
\begin{equation}
R^{kl}_{\tau,V/A}=\frac{1}{\Gamma_{e}} 
\int_{0}^{m_{\tau}^{2}}ds\,
w^{kl}(s)\,\frac{d\Gamma_{ud}^{V/A}}{ds},
\label{eq:rkldef}
\end{equation}
where $\Gamma_{e}$ is the electronic
width of the $\tau$ lepton, $d\Gamma_{ud}^{V/A}/ds$ is the
invariant mass distribution 
for the Cabbibo allowed semileptonic $\tau$ decays in the vector (V)
or axial-vector (A) channel and 
$w^{kl}(s)=(1-s/m_{\tau}^{2})^{k}(s/m_{\tau}^{2})^{l}$.  
 The QCD predictions for
$R^{kl}_{\tau,V/A}$ have the form:
\begin{eqnarray}
R_{\tau,V/A}^{kl}&=&\frac{3}{2}\,|V_{ud}|^{2}\,S_{EW}\,
R^{kl}_{0}(1+\delta^{kl}_{V/A}),\nonumber\\
\delta^{kl}_{V/A}&=&\delta^{kl}_{pt}+\delta^{kl}_{npt,V/A},
\label{eq:rklpred}
\end{eqnarray}
where $|V_{ud}|=0.9752$ and $S_{EW}=1.0194$. 
The $R^{kl}_{0}$ factor
denotes the parton model prediction.  The $\delta^{kl}_{pt}$ term
is the perturbative contribution, which is evaluated assuming 3 massless
quarks --- in this approximation it is universal for the V and A
channels. The $\delta^{kl}_{npt,V/A}$ term  
denotes the contribution from the nonperturbative QCD effects, which
are estimated using the SVZ approach~\cite{shif79}: 
\begin{equation}
\delta^{kl}_{npt,V/A}=\sum_{D=4,6...}\frac{1}{m^{D}_{\tau}}
\sum_{j} c^{kl}_{D,j}<O_{D,j}^{V/A}>,
\label{eq:svz}
\end{equation}
where $<O_{D,j}^{V/A}>$ are the vacuum expectation values of the gauge
invariant operators of dimension $D$ and $c^{kl}_{D,j}$ are
coefficients specific for the considered spectral moment and the type
of the operator.

The authors of \cite{exp,hoeck96} used $R^{00}_{\tau,V/A}$ and the
$R^{kl}_{\tau,V/A}$ 
moments with $k=1$ and $l=0,1,2,3$, and fitted the $D=4,6,8$
condensates. Since our goal is primarily to study the
theoretical uncertainties in the whole procedure, we shall adopt a
simplified approach, in which the nonperturbative contribution to
$R^{00}_{\tau,V/A}$ is approximated by the leading $D=6$ contribution,
coming from 
the four-quark condensates.
 (This means we neglect the $D=4$ contribution, which is
suppressed by an additional power of $\alpha_{s}$, the higher order
perturbative corrections to the coefficients of the $D=6$ operators,
and the $D\geq8$ contributions.)  We also use the $R^{12}_{\tau,V/A}$
moment, for which a similar approximation may be made, and for which
we have the relation 
$\delta^{12}_{(6)V/A}=-(70/13)\delta^{00}_{(6)V/A}$. 
This gives us two equations to fit two parameters:
$\Lambda^{(3)}_{\overline{MS}}$ (or, more conventionally, 
$\alpha_{s}^{\overline{MS}}(m_{\tau}^{2})$)  and one parameter
characterizing the nonperturbative contribution, which we take simply
to be 
$\delta^{00}_{(6)V/A}$.

The perturbative QCD corrections $\delta_{pt}^{kl}$ are evaluated
using a contour integral expression, which
relates them to the 
QCD correction $\delta_{\Pi}^{pt}$ to the so called Adler
function, i.e. the logarithmic derivative of the
transverse part of the vector/axial-vector current correlator:
\begin{equation}
\delta_{pt}^{kl}=\frac{i}{\pi}\int_{C}\frac{d\sigma}{\sigma}
f^{kl}\left(\frac{\sigma}{m_{\tau}^{2}}\right)\delta_{\Pi}^{pt}(-\sigma),
\label{eq:cont}
\end{equation}
where $f^{kl}(\sigma/m_{\tau}^{2})$ is a weight function specific
to the considered moment and $C$ in our case is assumed to be a circle
$\sigma=-m_{\tau}^{2}\exp(-i\theta)$, $\theta \in  [-\pi,\pi]$. 
The NNLO renormalization group improved perturbative expansion
for $\delta_{\Pi}^{pt}$ may be written in the form:
\begin{equation}
\delta_{\Pi}^{pt}(-\sigma) = a(-\sigma)[1+
r_{1}a(-\sigma)+r_{2}a^{2}(-\sigma)],
\label{eq:dpert}
\end{equation}
where $a=\alpha_{s}/\pi=g^{2}/(4 \pi^{2})$ denotes the running
coupling constant that satisfies the NNLO renormalization group equation:
\begin{equation}
\sigma \frac{d\,a}{d\sigma} = - \frac{b}{2}
\,a^{2}\,(1 + c_{1}a + c_{2}a^{2}\,).
\label{eq:rge}
\end{equation}
By evaluating numerically the integral (\ref{eq:cont}) with the
renormalization group improved expression for $\delta_{\Pi}^{pt}$
under the integral we resum to all orders some of the corrections,
which would appear in the ``naive'' expansion of $\delta_{pt}^{kl}$ in
powers of  $\alpha_{s}(m_{\tau}^{2})$ \cite{resum}. 

 The coefficients $r_{1}$, $r_{2}$ and $c_{2}$ are RS
dependent, but there exists a RS invariant combination:
\begin{equation}
\rho_{2}=c_{2}+r_{2}-c_{1}r_{1}-r_{1}^{2}.
\label{eq:rho2}
\end{equation}
For the $\delta_{\Pi}$ we have \cite{pert,chet97} $\rho_{2}=5.23783$.
We shall use $r_{1}$ and $c_{2}$ to parametrize the
freedom of choice of the RS in NNLO.

The dependence of $\delta_{pt}^{00}$ on the scheme parameters $r_{1}$
and $c_{2}$ was discussed in detail in \cite{racz96a} and the RS
dependence of $\delta_{pt}^{12}$ was investigated in
\cite{racz96c}. 
In both cases it was found that for
moderate values of 
$\Lambda^{(3)}_{\overline{MS}}$ the NNLO predictions have a
saddle point type of behavior as a function of $r_{1}$ and
$c_{2}$ and that the position of the saddle point is well
approximated by $r_{1}=0$ and $c_{2}=1.5\rho_{2}=7.857$.  
For very large
values of $\Lambda^{(3)}_{\overline{MS}}$ the RS-dependence
pattern is more complicated than a simple saddle point, but even
then the scheme parameters distinguished above belong to the
region of extremely small RS dependence. We shall therefore
accept these parameters as the PMS parameters in NNLO.

In order to obtain a quantitative estimate of the RS dependence of the
results we shall use the approach outlined in \cite{racz95}, based on the
existence of the RS invariant $\rho_{2}$. 
In \cite{racz95} it was proposed to calculate variation of the
predictions over the set of schemes for which the expansion
coefficients satisfy the condition:
\begin{equation}
\sigma_{2}(r_{1},r_{2},c_{2}) \leq l\,|\rho_{2}|,
\label{eq:condition}
\end{equation}
 where  
\begin{equation}
\sigma_{2}(r_{1},r_{2},c_{2})=|c_{2}|+|r_{2}|+c_{1}|r_{1}|+r_{1}^{2}.
\end{equation}
A motivation for the condition (\ref{eq:condition}) is that it
 eliminates schemes in which the expansions (\ref{eq:dpert}) and
 (\ref{eq:rge}) involve unnaturally large expansion coefficients, that
 introduce large cancellations in the expression for the RS
 invariant $\rho_{2}$. The constant $l\,$ in the condition
 (\ref{eq:condition}) controls the degree of cancellation that we
 want to allow in the expression for $\rho_{2}$. In our case we
 have for the PMS parameters $\sigma_{2}(\mbox{PMS})\approx2|\rho_{2}|$,
 so in order to take into account the schemes, which have the
 same --- or smaller --- degree of cancellation as the PMS scheme
 we take $l=2$. 

Let us begin with the fits in the vector channel. We use
the experimental values reported recently by ALEPH \cite{hoeck96}:
$R^{00}_{\tau,V}=1.782\pm0.018$,
$D^{12}_{\tau,V}=R^{12}_{\tau,V}/R^{00}_{\tau,V}=0.0532\pm0.0007$. 
For simplicity we neglect correlations between experimental errors for
$R^{00}_{\tau}$ and 
$D^{12}_{\tau}$.  
For comparison with other determinations of
$\alpha_{s}$ we evolve the fitted value of
$\alpha_{s}^{\overline{MS}}(m_{\tau}^{2})$ to the scale of $m_{Z}^{2}$
using the NNLO renormalization group equation and the NNLO matching
formula \cite{match} at $\mu=2m_{c},2m_{b}$. 

Using the PMS predictions we obtain from the NNLO fit in the vector
channel $\delta^{00}_{(6)V}=0.0156\pm0.0023$ and
$\Lambda^{(3)}_{\overline{MS}}=421\pm30\,\mbox{MeV}$, which
corresponds to $\alpha_{s}(m_{\tau}^{2})=0.356\pm0.017$ and
$\alpha_{s}(m_{Z}^{2})=0.1226\pm0.0018$. For comparison, using the
$\overline{\mbox{MS}}$ scheme we obtain
$\Lambda^{(3)}_{\overline{MS}}=441\pm32\,\mbox{MeV}$ and
$\delta^{00}_{(6)V}=0.0147\pm0.0025$, which corresponds to
$\alpha_{s}(m_{\tau}^{2})=0.367\pm0.018$ and
$\alpha_{s}(m_{Z}^{2})=0.1238\pm0.0018$.

\begin{figure}[htb]
 ~\epsfig{file=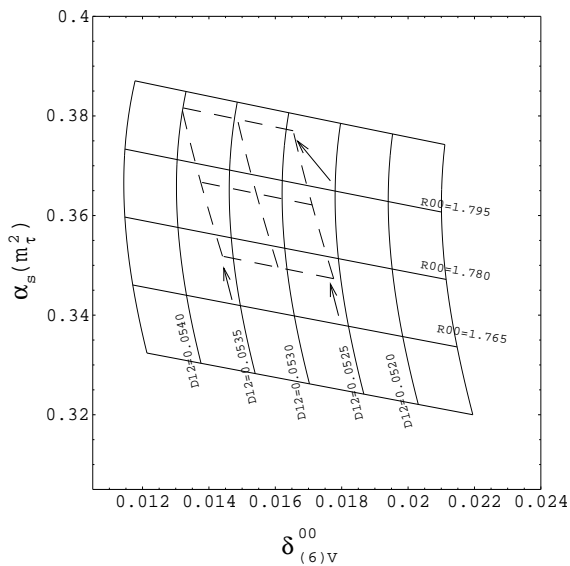,bbllx=0pt,bblly=0pt,bburx=212pt,bbury=211pt} 
\caption{Plot of the fitted values of $\alpha_{s}(m^{2}_{\tau})$
and $\delta^{00}_{(6)}$ in the vector channel as a function of
$R^{00}_{\tau,V}$ and $D^{12}_{\tau,V}$, obtained using the NNLO
PMS predictions. The dashed lines indicate the change in the plot
when the $\overline{\mbox{MS}}$ NNLO predictions are used
instead.}
\label{fig:ald6vec}
\end{figure}

In order to make our calculations  more generally useful  we show in
Fig.~\ref{fig:ald6vec} the results of the NNLO fit of
$\alpha_{s}(m^{2}_{\tau})$ and 
$\delta^{00}_{(6)V}$, obtained using the PMS predictions, 
as a  function of the experimental values of $R^{00}_{\tau,V}$ and
$D^{12}_{\tau,V}$. 

\begin{figure}[htb]
 ~\epsfig{file=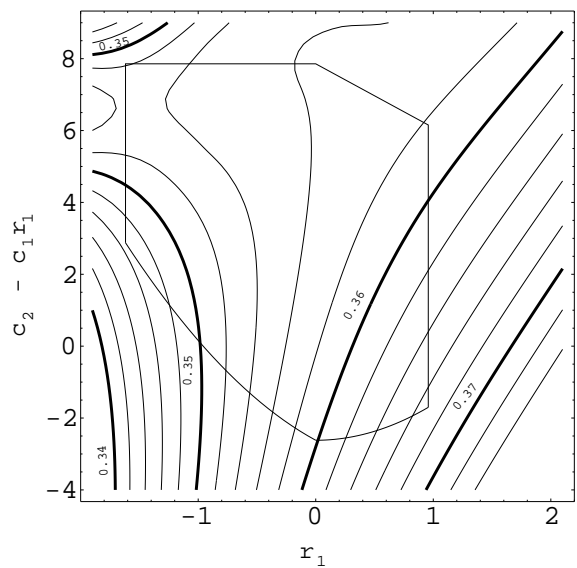,bbllx=0pt,bblly=0pt,bburx=212pt,bbury=211pt} 
\caption{The contour plot of the fitted value of
$\alpha_{s}(m^{2}_{\tau})$ in the vector channel as a function of
the RS parameters $r_{1}$ and $c_{2}$.  The region of the scheme
parameters satisfying the condition
(\protect{\ref{eq:condition}}) with $l=2$ has 
been also indicated.}
\label{fig:alfitvecrsdep}
\end{figure}

In Fig.~\ref{fig:alfitvecrsdep} we show how the value of
$\alpha_{s}(m_{\tau}^{2})$ resulting from the fit depends on the
parameters $r_{1}$ and $c_{2}$ specifying the renormalization
scheme in NNLO. By varying the RS parameters in the region satisfying
the condition~(\ref{eq:condition}) with $l=2$ we obtain the variation 
$0.347<\alpha_{s}(m^{2}_{\tau})<0.367$ 
($403\,\mbox{MeV}<\Lambda^{(3)}_{\overline{MS}}<440\,\mbox{MeV}$,
$0.1217<\alpha_{s}(m^{2}_{Z})<0.1238$). Analyzing in a similar way the
RS dependence of $\delta^{00}_{(6)V}$ we obtain
$0.0145<\delta^{00}_{(6)V}<0.0182$.

It is of some interest to perform the same fits using instead the
NLO predictions. (The PMS parameters in NLO are $r_{1}=-0.76$ for
$\delta^{00}_{pt}$ and $r_{1}=-0.64$ for $\delta^{12}_{pt}$.) Using the NLO
predictions in the PMS scheme we obtain 
$\delta^{00}_{(6)V}=0.0150$ and
$\Lambda^{(3)}_{\overline{MS}}=465\,\mbox{MeV}$. 
 Using the NLO
predictions in the 
$\overline{\mbox{MS}}$ scheme, we obtain
$\delta^{00}_{(6)V}=0.0148$ and
$\Lambda^{(3)}_{\overline{MS}}=527\,\mbox{MeV}$.

Similar fits may be performed in the axial-vector channel. We use
the experimental values reported recently by ALEPH \cite{hoeck96}:
 $R^{00}_{\tau,A}=1.711\pm0.019$,
$D^{12}_{\tau,A}=0.0639\pm0.0005$. Using the PMS scheme we obtain in
NNLO 
$\delta^{00}_{(6)A}=-0.0165\pm0.0018$ and
$\Lambda^{(3)}_{\overline{MS}}=380\pm34\,\mbox{MeV}$, which
corresponds to $\alpha_{s}(m_{\tau}^{2})=0.335\pm0.018$ and
$\alpha_{s}(m_{Z}^{2})=0.1203\pm0.0021$. For comparison, using the 
$\overline{\mbox{MS}}$ scheme we obtain 
$\delta^{00}_{(6)A}=-0.0168\pm0.0021$ and
$\Lambda^{(3)}_{\overline{MS}}=398\pm37\,\mbox{MeV}$, which
corresponds to $\alpha_{s}(m_{\tau}^{2})=0.344\pm0.019$ and
$\alpha_{s}(m_{Z}^{2})=0.1213\pm0.0021$.

Performing the variation of the RS parameters in the region 
satisfying the condition~(\ref{eq:condition}) with $l=2$ we
obtain  for the axial-vector channel $0.326<\alpha_{s}(m^{2}_{\tau})<0.343$
($364\,\mbox{MeV}<\Lambda^{(3)}_{\overline{MS}}<397\,\mbox{MeV}$,
$0.1193<\alpha_{s}(m^{2}_{Z})<0.1212$) and
$0.015<-\delta^{00}_{(6)A}<0.017$.

Summarizing, we find that in NNLO the
change from the $\overline{\mbox{MS}}$ scheme to the PMS scheme
results in the reduction of the fitted value of
$\alpha_{s}(m_{\tau}^{2})$ by approximately 0.01. 
($\alpha_{s}(m_{Z}^{2})$ is reduced by 0.001.) Also, the
difference between the NLO and NNLO results is much smaller in
the PMS scheme than in the $\overline{\mbox{MS}}$ scheme. The
extracted values of the QCD parameters appear to be relatively stable
with respect to change of the RS.
However, varying the scheme parameters $r_{1}$ and $c_{2}$ in
the region satisfying the condition~(\ref{eq:condition}) with $l=2$ we
obtain an uncertainty in the extracted value of $\alpha_{s}(m_{\tau}^{2})$
of approximately 0.02 (uncertainty in $\alpha_{s}(m_{Z}^{2})$ is 
0.002).

More details about the calculations and the results reported above may be
found in \cite{racz97a}.

\end{document}